\documentclass[conference]{IEEEtran}
\IEEEoverridecommandlockouts
\usepackage{tikz}
\nonstopmode
\usepackage{cite}
\usepackage{amsmath,amssymb,amsfonts}
\usepackage[linesnumbered,lined,ruled]{algorithm2e}
\usepackage{graphicx}
\usepackage{textcomp}
\usepackage{xcolor}
\usepackage{hyperref}
\usepackage{cleveref}
\usepackage{pdfpages}
\usepackage{commath}
\usepackage{optidef}
\usepackage[symbol]{footmisc}
\usepackage{lipsum}
\usepackage{enumitem}
\usepackage{svg}
\usepackage{subcaption}

\usepackage[left=1.62cm,right=1.62cm,top=1.7cm,bottom=2.67cm]{geometry}
\IEEEaftertitletext{\vspace{-0.5\baselineskip}}

\makeatletter
\newcommand{\nosemic}{\renewcommand{\@endalgocfline}{\relax}}
\newcommand{\dosemic}{\renewcommand{\@endalgocfline}{\algocf@endline}}

\let\oldnl\nl
\newcommand{\nonl}{\renewcommand{\nl}{\let\nl\oldnl}}
\makeatother

\def\BibTeX{{\rm B\kern-.05em{\sc i\kern-.025em b}\kern-.08em
    T\kern-.1667em\lower.7ex\hbox{E}\kern-.125emX}}

\begin{document}

\title{Optimum Power Allocation for Low Rank Wi-Fi Channels: A Comparison with Deep RL Framework}
\author{
    \IEEEauthorblockN{
         Muhammad Ahmed Mohsin\IEEEauthorrefmark{1}, Sagnik Bhattacharya\IEEEauthorrefmark{1}, Kamyar Rajabalifardi \IEEEauthorrefmark{1}, Rohan Pote\IEEEauthorrefmark{2}, John M. Cioffi\IEEEauthorrefmark{1}
        }
        
   \IEEEauthorblockA{\IEEEauthorrefmark{1}Dept. of Electrical Engineering, Stanford University, Stanford, CA, USA}

   \IEEEauthorblockA{\IEEEauthorrefmark{2}Samsung Semiconductors, San Diego, CA, USA}

    \IEEEauthorblockA{Email: \{muahmed, sagnikb, kfardi\}@stanford.edu, rohan.pote@samsung.com
    }
}

\maketitle

\begin{abstract}Upcoming Augmented Reality (AR) and Virtual Reality (VR) systems require high data rates ($\geq$ 500 Mbps) and low power consumption for seamless experience. With an increasing number of subscribing users, the total number of antennas across all transmitting users far exceeds the number of antennas at the access point (AP). This results in a low rank wireless channel, presenting  a bottleneck for uplink communication systems. The current uplink systems that use orthogonal multiple access (OMA) and the proposed non-orthogonal multiple access (NOMA), fail to achieve the required data rates / power consumption under predominantly low rank channel scenarios. This paper introduces an optimal power sub carrier allocation algorithm for multi-carrier NOMA, named minPMAC, and an associated time-sharing algorithm that adaptively changes successive interference cancellation decoding orders to maximize sum data rates in these low rank channels. This Lagrangian based optimization technique, although globally optimum, is prohibitive in terms of runtime, proving inefficient for real-time scenarios. Hence, we propose a novel near-optimal deep reinforcement learning-based energy sum optimization (DRL-minPMAC) which achieves real-time efficiency. Extensive experimental evaluations show that minPMAC achieves 28\% and 39\% higher data rates than NOMA and OMA baselines. Furthermore, the proposed DRL-minPMAC runs ~5 times faster than minPMAC and achieves 83\% of the global optimum data rates in real time.\footnote{This work was supported by research relationships between Stanford and Intel, Samsung and Ericsson.}
\end{abstract}


\section{Introduction}
The exponentially increasing number of upcoming real-time distributed wireless AR/VR systems necessitate lower latency, higher reliability, and significantly increased data rates. Mixed Reality (MR) applications require high data rates ($\geq 500$ Mbps) and low energy consumption ($\leq 100$ mW)~\cite{bhattacharya2024optimalpowerallocationtime}. With increasing number of subscribing users, the number of user transmitter antennas exceeds the number of access point (AP) antennas. This situation creates a rank deficient (low rank) wireless channel, which is a major bottleneck when it comes to satisfying the required data rates for every user. This work considers specifically an uplink multiple access channel (MAC), but can also be extended to downlink channels through the duality property \cite{book}.

Existing OMA power allocation methods that conform to latest Wi-Fi standards (802.11b/g/n/ac/be) deploy linear receivers, and thus, they do not satisfy these high-data-rate requirements for low-rank channel scenarios. This is because the OMA methods do not use nor decode the interference caused by other user's signal~\cite{yin2006ofdma}. Hence, this presents a challenge for WiFi-based AR/VR applications.

To address this limitation for OMA, non-orthogonal multiple access (NOMA) has been proposed~\cite{noma6, tse_viswanath_2005}. NOMA uses successive interference cancellation (SIC) to decode the signal at the AP. 
The decoding order, for SISO systems, is decided heuristically by the users' channel gains. NOMA demonstrates superior efficacy against OMA methods~\cite{6692652}. However, for MIMO scenarios~\cite{noma7}, there is no natural order given by the users' channel vectors. This leads to heuristic assumptions on SIC decoding order, e.g., based on the norm of the channel vector, etc. which lead to suboptimal resource allocation. Finding the optimal SIC decoding order for MIMO scenarios with multiple antennas at the AP and multiple users requires extensive optimization. We introduce an optimal power sub carrier allocation algorithm for multi-carrier NOMA, named minPMAC, and an associated time-sharing algorithm that adaptively changes successive interference cancellation decoding orders to maximize sum data rates in these low rank channels. However, due to longer runtimes, it is inefficient for real-time scenarios.

Deep reinforcement learning (DRL) has emerged as a popular optimization technique, particularly well-suited for networks with numerous constraints and optimization parameters, accelerating convergence. DRL-NOMA has been utilized in~\cite{wang2020drl, zhang2019energy} to optimize the uplink NOMA performance. It enhances the data rates while maintaining the energy efficiency.~\cite{wang2020drl}, considers subchannel assignment policy and then maximizes the weighted rate sum for the multiuser system.~\cite{zhang2019energy} explores a two-step model-free sub-carrier assignment and power-allocation algorithm to maximize energy efficiency. However, both these approaches are suboptimal because they consider heuristic-channel-gain based decoding order for their optimization problem. As per the author's best knowledge, there is no prior work achieving near optimal resource allocation in real time for low rank scenarios.

To summarize, the contributions of this work are:\\
\textbf{1.} We propose minPMAC - an optimal power-subcarrier allocation algorithm for multi-carrier NOMA, along with optimal decoding order, eliminating per-subcarrier decoding latency. Additionally, a dynamic time-sharing decoding order is introduced, allowing the system to switch between different decoding strategies to achieve higher data rates, thereby improving overall system efficiency without relying on any heuristic assumptions. \\
\textbf{2.} We propose DRL-minPMAC, which achieves 83\% of the global optimum sum data rates while being 5x faster than minPMAC (real-time). Dynamic decoding order for each time-step is considered in DRL-minPMAC as well for policy training and real time-inference. \\
\textbf{3.} Extensive experimental evaluations show that minPMAC achieves 28\% and 39\% higher data rates than NOMA and OMA baselines. The proposed DRL-minPMAC runs ~5 times faster than minPMAC and achieves 83\% of the global optimum data rates in real time. Furthermore, evaluation is tested for effectiveness across all scenarios against various DRL agents to benchmark consistency.

\section{System Model}

\subsection{System Description}
We consider an uplink MAC channel where 3 AR/VR users, equidistant from an AP with distance 3m each, transmit symbols to the receiver. We consider $N$ users, where $n_{th}$ user has $L_{x,n}$ antennas and transmit signal vector $\boldsymbol{x}_n \in \mathbb{R}^{L_{x,n}}$ to the AP with $L_y$ antennas. The received signal at the AP is given by the general equation:
\begin{equation} \label{eq:MIMO}
    \boldsymbol{y} = H \:.\: \boldsymbol{x} + \boldsymbol{n},
\end{equation}
where $H \in \mathbb{C}^{L_y \times L_x}$ denotes the channel matrix, and $\boldsymbol{n} \in \mathbb{R}^{L_y}$ is the additive white Gaussian noise (AWGN) at the receiver. Furthermore, $\boldsymbol{x} \in \mathbb{R}^{L_x}$ is the concatenation of all symbols $\boldsymbol{x}_n$ sent by the users to the receiver, and $\boldsymbol{y} \in \mathbb{R}^{L_y}$ is the received signal and $S$ is total subcarriers.\\
\textbf{NOMA SNRs} For the uplink NOMA transmission, all $N$ users are a single NOMA pair. The transmitted signal from the $N$ users at time step $t$ is $x_i[t] = \sum_{i=1}^{N} \sqrt{P_{i}} x_{i}[t]$, where $\sqrt{P_{i}}$ is the transmit power of user $n_i$ and $x_{i}[t]$ is the user signal. The received signal at the AP/BS is:
\begin{equation}
    y[t] = \sum_{i=1}^{N} H_{i}[t] x_{i}[t] + \sigma^2,
\end{equation}
where $H_{i}[t]$ is the channel for the user $n_i$ and $\sigma^2$ is the additive white Gaussian noise (AWGN).
 For SIC at AP/BS, we follow the heuristic channel gain based decoding order. The SINR equation based on SIC-decoding order for all users $n_i$ is:
 \begin{equation}
    \text{SINR}_{1} = \frac{P_{1} |H_{1}|^2}{P_{2} |H_{2}|^2 + P_{3} |H_{3}|^2 + \sigma^2}
\end{equation}
\begin{equation}
    \text{SINR}_{2} = \frac{P_{2} |H_{2}|^2}{P_{3} |H_{3}|^2 + \sigma^2}
\end{equation}
\begin{equation}
    \vdots
\end{equation}
\begin{equation}
    \text{SINR}_{N} = \frac{P_{N} |H_{N}|^2}{\sigma^2}
\end{equation}
where $SINR_{i}$ is the SINR for user $n_i$. The achievable data rate for the user $n_i$ is $R_{i} = \frac{B}{S} \log_2(1 + \text{SINR}_{i})$, where $B$ is the bandwidth and $S$ is the number of subcarriers.


\section{Optimum Power Subcarrier Allocation using minPMAC}
\begin{enumerate}[wide=\parindent]
    \item 
\textit{\textbf{minPMAC problem formulation:}} MinPMAC's primary objective is to maximize the sum-rate achieved given the minimization of energy (or power) over $S$ subcarriers. The optimization formulates a primal dual optimization problem, where the primal problem deals with minimization of energy subject to minimum data rates. The converse problem maximizes the sum-rate subject to energy consumption constraint.
\begin{equation}
\begin{aligned}
&\min_{\left\{Q_{\boldsymbol{s} \boldsymbol{s}}(i, j)\right\}} \sum_{i=1}^{N} \sum_{j=1}^{S} \alpha_i \cdot \operatorname{tr}\left(Q_{\boldsymbol{s} \boldsymbol{s}}(i, j)\right) \\
&\text{D1}: \mathbf{r} = \sum_{j=1}^{S} \left[ r_{1, j}, r_{2, j}, \ldots, r_{N, j} \right]^T \succeq \mathbf{r}_{\text{th}} \succeq \mathbf{0} \\
&\text{D2}: \sum_{i \in \Omega}\sum_{j=1}^{S} r_{i, j} \leq \log_2 \left| \frac{\mathbf{Z}_{\text{int}} + \sum_{i \in \Omega} \mathbf{G}_{i, j} Q_{\boldsymbol{s}\boldsymbol{s}}(i, j) \mathbf{G}_{i, j}^H}{\mathbf{Z}_{\text{int}}} \right|, \\
&\quad \forall \Omega \subseteq \{1, 2, \dots, N\} \\
&\text{D3}: \mathbf{Q}_{\boldsymbol{s}\boldsymbol{s}}(i,j) \succcurlyeq \mathbf{0}, \quad \forall i \in \{1, 2, \dots, N\}, \forall j \in \{1, 2, \dots, S\}
\end{aligned}
\label{eq: formulation}
\end{equation}
$\mathbf{Q}{\boldsymbol{ss}}(i,j) \in \mathbb{R}^{M{s,i}\times M_{s,i}}$ represents the auto-correlation matrix of signal $\boldsymbol{s}$ for user $i$ on the $j^{th}$ subcarrier, while $\alpha_i$ is a weight prioritizing the power contribution of user $i$. The term $r_{i,j}$ denotes the data rate for user $i$ on subcarrier $j$, and $\mathbf{r}_{\text{th}}$ represents the required data rate thresholds. D1 enforces the threshold requirement, D2 ensures that the rate sum for any user subset is within the theoretical achievable region, and D3 ensures the covariance matrix is positive semi-definite. 
\item
\textit{\textbf{minPMAC Algorithm:}}
MinPMAC aims to determine the optimal power allocation  across all subcarriers in order to achieve the target data rates. The optimization is independent of the SIC decoding order, and decoupling power and subcarrier allocation maintains problem convexity
\item 
\textit{\textbf{Canonical SIC:}}
The receiver employs successive interference cancellation (SIC) for detecting the transmitted symbols. This choice is motivated by the fact that, for a given user power allocation, a canonical SIC decoding order can achieve the information-theoretic sum-rate capacity bound, provided that the optimal decoding order is selected \cite{book}. Let us define the decoding order vector as $\boldsymbol{\phi}(.)$, where $\boldsymbol{r_\phi}(1)$ represents the decoding order for user 1, and so on. Conversely, we define $\boldsymbol{r_\phi}^{-1}(.)$ as the inverse function of $\boldsymbol{r_\phi}(.)$, with $\boldsymbol{r_\phi}^{-1}(1)$ indicating the user index that is decoded first. Based on this notation, we can express the achieved data rate of the user $\boldsymbol{r_\phi}^{-1}(i)$ as \cite{book}:
\begin{equation}\label{eq:SIC}
\begin{aligned}
r_{\boldsymbol{\phi}^{-1}(i)} &= \sum_{k=i}^N r_{\boldsymbol{\phi}^{-1}(k)} - \sum_{k=i+1}^N r_{\boldsymbol{\phi}^{-1}(k)} \\
& = \footnotesize{ \log_2 \left( \left| \frac{\mathbf{Z}_{\text{int}} + \sum_{k=i}^N \mathbf{G}_{\boldsymbol{\phi}^{-1}(k)} 
Q_{\boldsymbol{s}\boldsymbol{s}}(\boldsymbol{\phi}^{-1}(k)) \mathbf{G}^H_{\boldsymbol{\phi}^{-1}(k)}}
{\mathbf{Z}_{\text{int}} + \sum_{k=i+1}^N \mathbf{G}_{\boldsymbol{\phi}^{-1}(k)} 
Q_{\boldsymbol{s}\boldsymbol{s}}(\boldsymbol{\phi}^{-1}(k)) \mathbf{G}^H_{\boldsymbol{\phi}^{-1}(k)}} \right| \right)}
\end{aligned}
\end{equation}
where $\mathbf{Z}_{\text{int}}$ is the noise auto-correlation matrix, and $\mathbf{G}_{\boldsymbol{\phi}^{-1}(i)}$ denotes the channel matrix between the base station and user $\boldsymbol{\phi}^{-1}(i)$. Equation~\eqref{eq:SIC} is derived by considering the SIC assumption, where each user subtracts the interference caused by the previously decoded users' symbols, and treats the remaining users' signals as noise. Deriving the optimal decoding order $\boldsymbol{\phi}$ among $N!$ possible orders is non-trivial for vector channels without heuristic assumptions.
\item 
\textit{\textbf{Optimal Decoding Order Derivation for minPMAC: }}MinPMAC finds the optimal decoding order across all subcarriers. A heuristic decoding order based on channel state information (CSI) leads to suboptimal power allocation solutions. To determine the optimal decoding order, the method starts by solving the dual problem of Equation~\ref{eq: formulation} as:
\begin{equation}
\begin{aligned}
& \max_{\left\{ Q_{\boldsymbol{s} \boldsymbol{s}}(i, j) \right\}} \sum_{i=1}^{N} \theta_i \cdot \left( \sum_{j=1}^{S} r_{i, j} \right) \\
& \text{D1}: \mathcal{E} = \sum_{j=1}^{S} \left[ \mathcal{E}_{1,j}, \mathcal{E}_{2,j}, \ldots, \mathcal{E}_{N,j} \right]^T \preceq \boldsymbol{\mathcal{E}}_{\max} \\
& \text{D2}: \sum_{i \in \Omega} \sum_{j=1}^{S} r_{i,j} \leq \log_2 \left| \mathbf{Z}_{\text{int}} + \sum_{i \in \Omega} \mathbf{G}_{i,j} \cdot Q_{\boldsymbol{s}\boldsymbol{s}}(i, j) \cdot \mathbf{G}_{i,j}^H \right| \\
& \quad \qquad - \log_2 \left| \mathbf{Z}_{\text{int}} \right| \quad \forall \Omega \subseteq \{1, 2, \cdots, N\} \\
& \text{D3}: Q_{\boldsymbol{s}\boldsymbol{s}}(i, j) \succcurlyeq \mathbf{0} \quad \forall i \in \{1, 2, \ldots, N\}, \forall j \in \{1, 2, \ldots, S\}
\end{aligned}
\end{equation} 
where $\mathcal{E}_{i,j} = \operatorname{tr}(Q_{\boldsymbol{s}\boldsymbol{s}}(i,j))$ represents the energy allocated to user $i$ on subcarrier $j$, while $\theta_i$ are the non-negative weights assigned to user data rates. Constraints D1 ensure that the total energy remains within limits, D2 enforces the data rate capacity condition, and D3 guarantees that the covariance matrices are positive semi-definite.
As demonstrated in \cite{book}, the maximum achievable weighted sum rate can always be attained. It has also been shown that for this optimal sum rate, the decoding order $\mathbf{\phi}$ must satisfy the following inequality, which depends on the dual variables $\theta_{n}$~\cite{book}:

\begin{equation}
\label{eq:decoding_order}
\theta_{\mathbf{\phi}^{-1}(N)} \geq \theta_{\mathbf{\phi}^{-1}(N-1)} \geq \dots \geq \theta_{\mathbf{\phi}^{-1}(1)}
\end{equation}
Thus, the optimal decoding order is determined by the ranking of the dual variables associated with the minimum required data rates in the energy minimization problem. Solving this convex optimization problem using a primal-dual method, as described in Equation~\eqref{eq: formulation}, yields both the dual variables $\theta$ and the corresponding optimal decoding order.
\item 
\textit{\textbf{Time Sharing:}} To ensure that the required data rates are met, it is crucial to have a well-defined decoding order based on a strict ranking of the \(\theta_{n}\) values for all users \(n \in \{1, 2, \ldots, N\}\). However, Equation~\ref{eq:decoding_order} does not always result in strict inequalities, which can lead to scenarios where equal \(\theta_{n}\) values make it impossible to find a unique decoding order that satisfies the target data rates~\cite{book}. In such cases, conventional NOMA techniques that rely on a single heuristic decoding order may not be effective. To overcome this challenge, the proposed method introduces a novel approach that utilizes time-sharing to combine multiple decoding orders when a single decoding order is not sufficient to achieve the desired data rates. This time-sharing algorithm addresses situations where \(\theta_{n}\) values are the same by grouping users with identical \(\theta_{n}\) into clusters. Within each cluster, all possible user order permutations are considered. 
MinPMAC groups users with identical \(\theta_{i}\) values into clusters, treating each cluster as a single composite user. These clusters, along with the remaining individual users, are then arranged in a strict order based on their \(\theta_{i}\) values. By considering all possible permutations within each cluster, the minPMAC algorithm generates a set of candidate decoding orders, denoted as \(N_{o}\), all of which require the same power allocation. The time-sharing algorithm evaluates each decoding order using successive interference cancellation (SIC) with optimal power allocation, represented by \(R_{j}\). It then determines a linear combination of these rates using linear programming to satisfy the target data rates. The time-sharing weights, denoted as \(w_{j}\), indicate the fraction of time each decoding order is utilized, ensuring that the target rates are achieved even when a single decoding order is insufficient. The linear programming formulation is:
\begin{equation}
\begin{aligned}
\text{minimize} \quad & \sum_{j=1}^{N_{o}} y_{j} \\
\text{subject to} \quad & \text{D1: } y_{j} \in\{0,1\} \quad \forall j \in\left\{1, \ldots, N_{o}\right\} \\
& \text{D2: } w_{j} \leq y_{j} \quad \forall j \in\left\{1, \ldots, N_{o}\right\} \\
& \text{D3: } w_{j} \geq 0 \quad \forall j \in\left\{1, \ldots, N_{o}\right\} \\
& \text{D4: } \sum_{j=1}^{N_{o}} w_{j} = 1 , \text{D5: } \sum_{j=1}^{N_{o}} R_{j} \cdot w_{j} = \mathbf{r}_{\text{th}}
\end{aligned}
\end{equation}

This formulation introduces a binary vector \(y\) of length \(N_{o}\). The objective function and constraint D2 work together to minimize the number of non-zero time-sharing weights, effectively reducing the number of decoding orders used in the time-sharing scheme. This is important because switching between decoding orders during time-sharing introduces latency, and minimizing the number of switches helps to reduce this latency. Constraints D3 and D4 ensure that the time-sharing weights are non-negative and sum to 1, as they represent the proportion of time each decoding order is applied. Finally, constraint D5 guarantees that the time-shared average data rates are equal to the target user data rates, denoted as \(\mathbf{r}_{\text{th}}\).

\end{enumerate}

\section{Near Optimal Power Subcarrier Allocation using DRl-minPMAC}
\begin{enumerate}[wide=\parindent]
    \item 
\textit{\textbf{DRL-minPMAC problem formulation:}} For the DRL-minPMAC problem formulation, the primary objective maximizes energy efficiency over $S$ subcarriers. The DRL-minPMAC algorithm jointly minimizes the transmit power for all three users over $S$ subcarriers and each time step $t$, denoted as, $\mathbf{P} \triangleq {p_i, \forall i}$ and simultaneously maximizing the sum-rate for all users. Mathematically, the optimization is:
    
\begin{maxi!}|s|
{\mathbf{P_i}} 
{\sum_{t \in \mathcal{T}} \frac{R_{\text{sum}}[t]}{P_{\text{total}}[t]}}{\label{eq:problem_ee}}{} 
\addConstraint{P_i \leq P_t,~\forall i \in \mathcal{N}}{\label{eq:power_csrt}} 
\addConstraint{R_i \geq R_{\text{min}},~\forall i \in \mathcal{N}}{\label{eq:rate_csrt}} 
\addConstraint{P_{\text{total}}[t] \leq P_{\text{max}},~\forall t \in \mathcal{T,}}{\label{eq:total_power_csrt}} 
\end{maxi!}
where $R_{\text{sum}}[t]$ is the sum rate at time step $t$, $P_{\text{total}}[t]$ is the total power consumption at time step $t$, $P_t$ is the constant maximum transmit power for each user, $R_{\text{min}}$ is the minimum required rate for each user, and $P_{\text{max}}$ is the maximum total power constraint.
\item
\textit{\textbf{MDP Formulation for DRL-minPMAC:}}
Following the problem formulation in [\ref{eq:problem_ee}], DRL-minPMAC reformulates this as a single-agent Markov Decision Process (MDP) that operates in discrete time steps. The agent's action is determined solely by the current state. The MDP is defined by the tuple [$\mathcal{S}$, $\mathcal{R}$, $\mathcal{A}$, $\mathcal{P}$, $\gamma$], where $\mathcal{S}$ denotes the state space of potential environment states, $\mathcal{R}$ represents the reward function for policy learning, $\mathcal{P}$ describes the state transition probabilities, and $\gamma$ is the discount factor applied to future rewards and exploration. At each time step $t$, the agent observes the current state $s_t$, selects an action $a_t$ based on its policy, transitions to a new state $s_{t+1}$, and receives a reward $\mathcal{R}(s_t, a_t)$. The explanation of state space, action space, reward function and the DRL agent is:
\begin{enumerate}[wide=\parindent]
    \item \textit{State Space $\mathcal{S}$:} 
The environment's state at time step $t$ consists of  achievable rates $\mathbf{\text{$R_i$}}[t]$, the total power $\mathbf{\text{$P_t$}}[t]$, and each user's individual power over all subcarriers, $\mathbf{\text{$P_i$}}[t]$. Formally, the state space is:
        \begin{equation}
              s_t = \{\mathbf{\text{$R_i$}}[t], \mathbf{\text{$P_t$}}[t], \mathbf{\text{$P_i$}}[t]\} \in \mathbb{R}^{\text{dim}_\mathcal{S}}.
        \end{equation}
where $\text{dim}_\mathcal{S}= \sum_{i \in \mathcal{N}} N_i+1+(N_i.S) $ is the dimension of the state space, where $N_i\triangleq$ the number of users.

    \item \textit{Action Space $\mathcal{A}$:} The actions available for the agent to learn the policy function consists of transmit powers $P_i$ for each user, $N_i$, $\forall\, \mathbf{S}$. Specifically, the action space at time step $t$ is:
    \begin{equation}
              a_t = \{\mathbf{\text{$P_t$}}[t], \mathbf{\text{$P_i$}}[t]\} \in \mathbb{R}^{\text{dim}_\mathcal{A}}.
        \end{equation}
          where $\text{dim}_\mathcal{A}=\sum_{i \in \mathcal{N}} 1 + (N_i.S) $ is the dimension of the action space.

    \item \textit{Reward Function $\mathcal{R}$:} The reward function is pivotal for defining an RL agent's learning trajectory. The choice here optimizes multiple objectives, including rate achievement, power efficiency, fairness among users, and power limit adherence. At each time step $t$, the reward is: 
\begin{equation}
\mathcal{R}(s_t, a_t) = R_{rate} + R_{eff} - P_{power} + R_{fair}
\end{equation}
where, 
\begin{equation}
R_{rate} \triangleq w_{1,i} \cdot \tanh\left(\frac{\sum_{i=1}^{N} R_i}{R_{min}} - 1\right)
\end{equation}
where $R_i$ represents the rate of user $i$, and $R_{min}$ is the minimum required rate for system performance. Also,
\begin{equation}
R_{eff} \triangleq w_{2,i} \cdot \tanh\left(\frac{\sum_{i=1}^{N} R_i}{\sum_{i=1}^{N} P_i + \epsilon} \cdot 10^{-6}\right)
\end{equation}
where $P_i$ is the power consumption of user $i$. To ensure the power constraints are respected:
\begin{equation}
P_{power} \triangleq w_{3,i} \cdot \max\left(0, \frac{\sum_{i=1}^{N} P_i}{P_{total}} - 0.8\right)
\end{equation}
where $P_{total}$ is the total power limit across all users. This penalty is applied only when the total power exceeds 80\% of the allowed limit. Fairness among users is promoted by rewarding the system when rates are evenly distributed. The fairness reward is calculated as:
\begin{equation}
R_{fair} \triangleq w_{4,i} \cdot \left(1 - \frac{\sum_{i=1}^{N} |R_i - \bar{R}|}{\bar{R} \cdot N + \epsilon}\right)
\end{equation}
where $\bar{R} = \frac{1}{N} \sum_{i=1}^{N} R_i$ is the average rate across all $N$ users. Furthermore, the  weights $\mathbf{W}[t] = \{w_{1,i}, w_{2,i}, w_{3,i}, w_{4,i}, \forall i, T\}$ are trainable weights whose values change as the training progresses or can be sight after trial and error.
\end{enumerate}

    \item \textit{\textbf{PPO Algorithm:}} 
PPO algorithm focuses on discrete action space for the DRL-minPMAC algorithm. The PPO framework consists of an actor network responsible for generating discrete actions $a_r$. The first few layers of the actor network are used for feature extraction and state information encoding. Furthermore, a single critic network is employed to estimate the value function $V(s_t)$, which is used to compute a variance-reduced advantage function estimate $\hat{A}_t$ for policy optimization. Following the implementation approach outlined in~\cite{mnih2016asynchronous}, the policy is executed for a fixed duration of $\hat{T}$ time steps. During this execution, the action $\hat{A}_t$ at each time step $t$ is calculated:
\begin{equation}
    \label{eq:adv}
    \hat{A}_t = \sum_{k=0}^{\hat{T}-1} \gamma^k r_{t+k} + \gamma^{\hat{T}} V(s_{t+\hat{T}}) - V(s_t),
\end{equation}
where $\hat{T}$ is much smaller than the length of the episode $T$.

To derive the stochastic policy \(\pi_{\theta_d}(a_t|s_t)\) for discrete actions, the actor network produces \(|a_{\text{R}}|\) logits, which are subsequently processed through a \(\mathrm{softmax}\) function to yield a probability distribution across the feasible discrete actions. For discrete actions, the objective function is expressed as
\begin{equation}
    \label{eq:clip}
    L_d^{\text{CLIP}}(\theta_d) = \hat{\mathbb{E}}_t \left[\min\left(r_t^d(\theta_d) \hat{A}_t, \Im(r_t^d, \theta_d, \epsilon) \hat{A}_t\right)\right],
\end{equation}
where \(\Im(r_t^d, \theta_d, \epsilon) = \text{clip}(r_t^d(\theta_d), 1 - \epsilon, 1 + \epsilon)\), \(r_t^d(\theta_d) = \frac{\pi_{\theta_d}(a_t|s_t)}{\pi_{\theta_d}^{\text{old}}(a_t|s_t)}\) is the importance sampling ratio, and \(\epsilon\) represents the clipping parameter. \( \theta_d \) denotes the set of parameters (weights and biases) that define this discrete actor network.

\end{enumerate}

\begin{table}[b!]
    \centering
    \caption{Simulation Parameters for PPO Algorithm}
    \label{tab:params}
    \resizebox{0.99\columnwidth}{!}{%
        \begin{tabular}{|c|c|c|c|}
            \hline
            \textbf{Parameter}                      & \textbf{Value} & \textbf{Parameter}            & \textbf{Value}   \\
            \hline
            \hline
            Bandwidth $B$   & $80$ MHz   & Learning rate                 & $5\text{e}-4$ \\
            Frequency $f$ & $2.49$GHz   & Clipping parameter $\epsilon$ & $0.2$            \\
            Penalty constant $K_{\text{viol}}$     & $50$            & Discount factor $\gamma$      & $0.98$           \\
            Minimum distance $d_{\text{min}}$       & $3$ m         & Number of episodes $N$        & $1000$            \\
            Time slots per episode $T$              & $1000$          & Number of epochs $E$          & $100$             \\
            Number of neurons                       & $64$           & Batch size $B$                & $128$            \\
            \hline
        \end{tabular}%

    }
\end{table}

\begin{table}[b!]
\centering
\caption{Received SNR [dB] vs. Achieved Data rate [Mbps] per User for PPO (No Fairness)}
\label{Sumrate}
\resizebox{0.99\columnwidth}{!}{%
\begin{tabular}{|c|c|c|c|c|c|c|}
\hline
\textbf{Data rate (Mbps)} & \textbf{-50} &  \textbf{-30} & \textbf{-10}  & \textbf{20}  \\
\hline
\hline
\textbf{minPMAC} & [0.036, 0.036, 0.036] & [2.795, 2.781, 2.798] & [95.224, 79.414, 65.610] & [676.891, 548.964, 379.887]  \\
\textbf{DRL-minPMAC} & [0.078, 0.023, 0.006]  & [6.195, 2.206, 0.009] & [170.427, 29.175, 0.745]  & [517.262, 477.725, 403.048]\\
\textbf{Brute Force} & [0.080, 0.016, 0.012]  & [5.671, 2.794, 0.075]  & [181.899, 22.346, 6.555] & [523.463, 511.56, 415.357] \\
\textbf{OMA} & [0.036, 0.036, 0.036]  & [2.789, 2.778, 2.798] & [65.751, 65.507, 65.604] & [379.732, 379.181, 381.236]  \\
\hline
\end{tabular}%
}
\end{table}

\begin{figure*}[t!]
     \centering
     \begin{subfigure}[t]{0.66\columnwidth} 
         \centering
         \includegraphics[width=\textwidth]{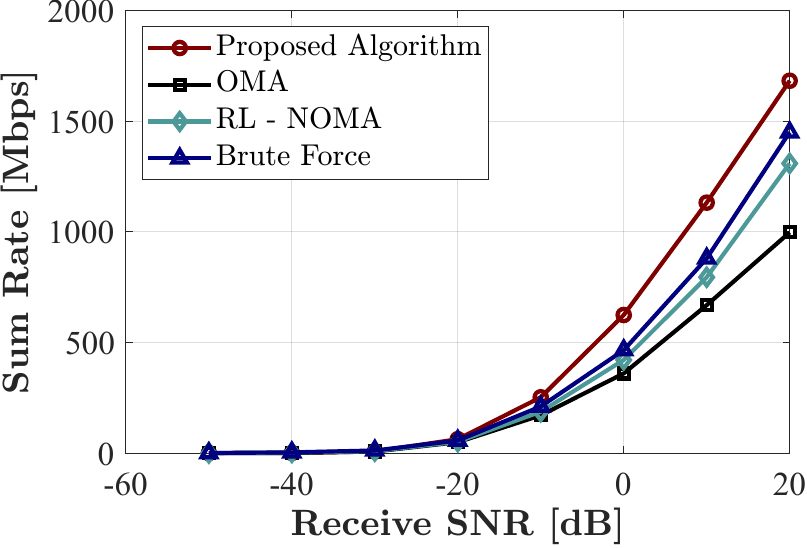}
         \caption{Sum Rate vs. receive SNR for Single Antenna Per User with distances from AP=\{3m, 3m, 3m\}.}
         \label{fig:Sumrate}
     \end{subfigure}
     \hspace{-0.01\columnwidth} 
     \begin{subfigure}[t]{0.66\columnwidth} 
         \centering
         \includegraphics[width=\textwidth]{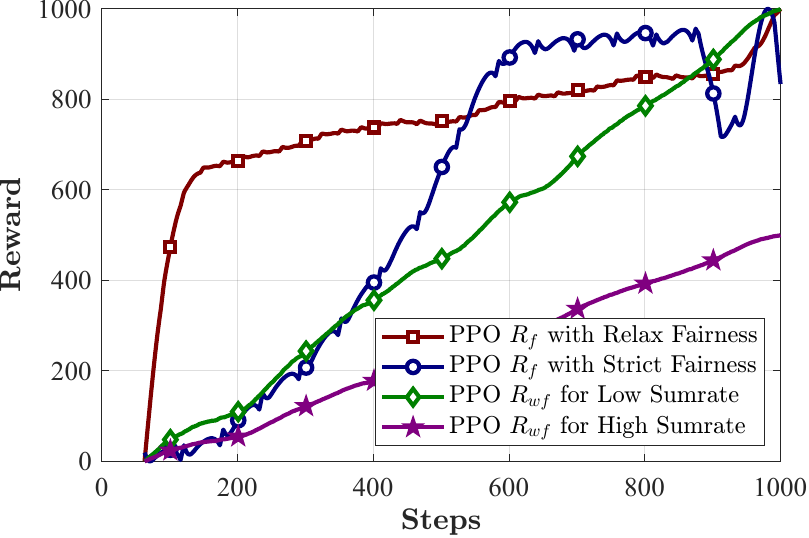}
         \caption{Time steps vs. reward function with different RL policies with normalized range.}
         \label{fig:reward}
     \end{subfigure}
     \hspace{-0.01\columnwidth} 
     \begin{subfigure}[t]{0.66\columnwidth} 
         \centering
         \includegraphics[width=\textwidth]{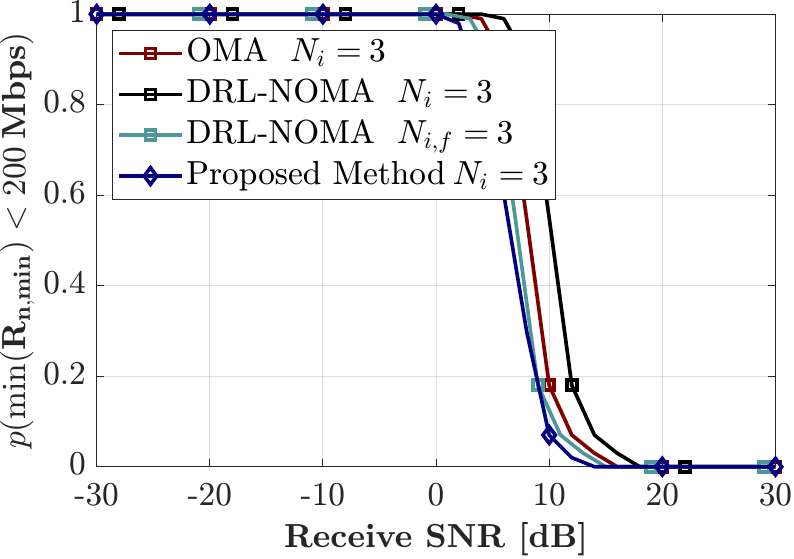} 
         \caption{Receive SNR vs. outage probability for Single Antenna Per User with distances from AP=\{3m, 3m, 3m\}.}
         \label{fig:outage}
     \end{subfigure}
     \caption{\textbf{(a)} Sum rate: minPMAC, DRL-minPMAC vs. OMA, Brute Force, \textbf{(b)} Reward vs. time: convergence behaviors and \textbf{(c)} Outage: DRL-minPMAC (fairness) vs. OMA.}
     \label{fig:spectral_efficiency}
\end{figure*}
\begin{figure*}[t!]
     \centering
     \begin{subfigure}[t]{0.66\columnwidth} 
         \centering
         \includegraphics[width=\textwidth]{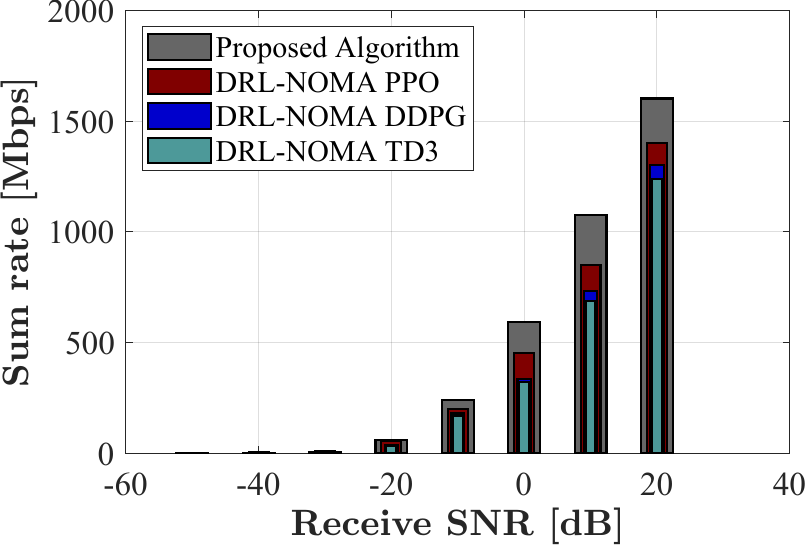}
         \caption{Receive SNR vs. sum rate for single antenna per user against baseline  DRL-NOMA agents.}
         \label{fig:comparison}
     \end{subfigure}
     \hspace{-0.01\columnwidth} 
     \begin{subfigure}[t]{0.66\columnwidth} 
    
         \centering
         \includegraphics[width=\textwidth]{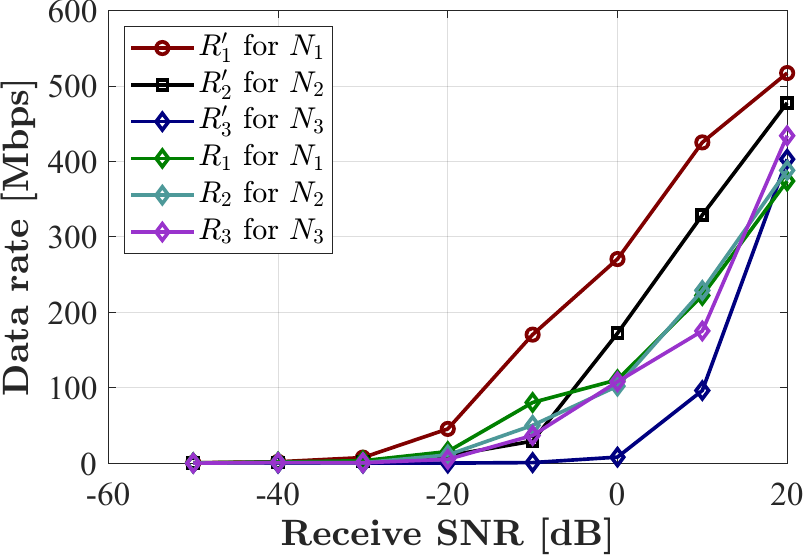}
         \caption{Receive SNR vs. data rate for single antenna per user (w and w/o fairness)}
         \label{fig:fairness_datarate}
     \end{subfigure}
     \hspace{-0.01\columnwidth} 
     \begin{subfigure}[t]{0.66\columnwidth} 
         \centering
         \includegraphics[width=\textwidth]{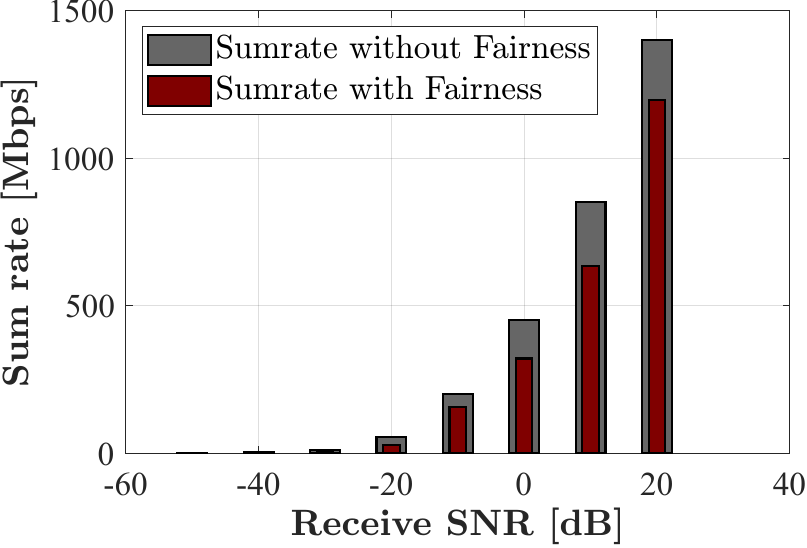}
         \caption{Receive SNR vs. sum rate for single antenna per user (w and w/o fairness).}
         \label{fig:Sumrate_fairness}
     \end{subfigure}
     \caption{\textbf{(a)} Sum rate: DRL-minPMAC (PPO, DDPG, TD3); \textbf{(b)} Data rate/user and sum rate/user w/wo fairness reward (DRL-minPMAC).}
     \label{fig:spectral_efficiency}
\end{figure*}

\section{Complexity and Convergence Analysis}
The complexity of the DRL-minPMAC algorithm is measured in terms of multiplications per iteration, which depends on the number of parameters in the policy and value networks. The overall complexity can be expressed as:

\begin{equation}
    \mathcal{O} \left( \frac{\alpha \beta \gamma}{\delta} \left( \sum_{q=1}^{Q_\pi} n_q n_{q-1} + \sum_{q=1}^{Q_v} n_q n_{q-1} \right) \right)
\end{equation}
where $\alpha$ is the number of epochs, $\beta$ the steps, $\gamma$ the number of networks, and $\delta$ the batch size. Assuming a uniform number of neurons $n$ in each hidden layer, and $d_{\text{in}}$, $d_{\text{out}}$ as the input and output dimensions, it simplifies to:

\begin{equation}
    \mathcal{O} \left( \frac{\alpha \beta \gamma}{\delta} \left( 2 d_{\text{in}} n + (Q_\pi + Q_v - 4) n^2 + n (d_{\text{out}} + 1) \right) \right)
\end{equation}
The convergence of PPO is difficult to analyze~\cite{challita2019interference}, since neural networks are highly dependent on the choice of hyperparameters. MinPMAC algorithm is optimized using the CVX package, which utilizes interior-point methods for optimization. This choice of solver is particularly effective for our problem structure, as it can handle the non-linear constraints efficiently. Moreover, our optimization problem can be formulated as a semi-definite programming (SDP) problem, which allows us to leverage powerful convex optimization techniques. The computational complexity of minPMAC is primarily determined by the SDP solver. For an SDP with n variables and m constraint matrices of size p × p, the worst-case complexity is of the order $\mathcal{O}(n^2m^{2.5} + m^{3.5})$ per iteration of the interior-point method. In our specific formulation, this translates to a complexity of $\mathcal{O}(N^6 \log(1/\epsilon))$, where N is the problem size (related to the number of users and frequency tones in our system) and $\epsilon$ is the desired accuracy.

\section{Numerical Results}


\subsection{Simulation Setup}
In our experiments, we use the WINNER\_Indoor\_A1 channel model, which is suitable for typical indoor deployments, such as Wi-Fi. In addition, we consider the scenario where the number of users ($N$) exceeds the number of antennas at AP ($L_y$). Users are positioned equidistantly from the AP and placed close to each other to increase the cross-talk effect. This setup allows our proposed algorithm to exploit the higher interference levels for improved performance. For simulations, we generated 1,000 channel samples using QuaDRiGa in MATLAB. We compare our results with baseline methods of OMA, DRL-NOMA and brute force.

\subsection{Results}
In Fig.~\ref{fig:Sumrate} shows the sum-rate for 3 users located at equal distance of 3 m from the AP against varying receive SNR at the AP. There is a substantial improvement in the currently deployed OMA method from the proposed minPMAC algorithm. On average, there is a 39\%, 15\%, 16\%,  improvement of the proposed minPMAC algorithm from the baseline OMA, Brute Force and DRL-minPMAC methods. For DRL-minPMAC and Brute force, the cross talk between users is utilized but due to non-optimal decoding order they also do not reach optimum performance and OMA doesn't utilize cross-talk at all. 

Fig.~\ref{fig:reward}, shows the average cumulative reward achieved by PPO for DRL-minPMAC algorithm against time steps. The figure demonstrates an increasing reward function, meaning that the agent is learning and exploring the environment as time steps increase. The increasing behavior demonstrates successful acquisition of an effective policy for the objective function

Fig~\ref{fig:outage}, shows the outage probability of 3 users, for the worst user having data rate of less than 200 Mbps. We observe that minPMAC outperforms in terms of outage as well (GDFE) against DRL-minPMAC, OMA and NOMA. DRL-minPMAC for $N_i$ = 3 has the highest outage probability because it does not consider fairness and hence, to achieve maximum sum rate, most power is allocated to stronger users. $N_{if}$ = 3 has better outage because we ensure fairness among the users and hence it performs better than existing OMA technique.

Fig~\ref{fig:comparison}, demonstrates the comparison of sum-rate achieved against varying receive SNR at the AP for different DRL-minPMAC agents against the minPMAC. The best performing DRL agent is PPO due to its clipped subjective surrogate function. It offers stability and sampling efficiency. DDPG struggles with complex environments due to its sensitivity to hyperparameters. TD3, despite being more stable, takes more time to converge. MinPMAC, performs better than all DRL agents, especially for higher SNRs.

Fig~\ref{fig:fairness_datarate} shows the comparison of the individual data rates of the users under fairness and unfairness conditions for the DRL-minPMAC. To introduce fairness in the system model, this result adds a reward for fairness and penalty if high data rate deviations among users. We observe that $R_{i}^{\prime}$ depicts data rate without fairness reward and $R_i$ shows the data rate of individual users with fairness reward. Fig~\ref{fig:Sumrate_fairness} observe that overall sum rate decreases with increasing fairness because more energy is required to meet the same rate for worse channel conditions.


\section{Conclusion and Future Work}
The proposed minPMAC and DRL-minPAC notably enhance the energy efficiency compared to baseline NOMA and current in use OMA methods. The results become prominent in rank deficient conditions, where the proposed minPMAC algorithm performs better for all SNRs. Our proposed methodology achieves high data rates under low SNRs proving optimum for VR/AR applications. MinPMAC outperforms NOMA and OMA by 28\% and 39\% respectively. Furthermore, DRL-minPMAC being near optimal and 5x faster than minPMAC based on Lagrangian optimization and is more efficient for real time inference and application. For future work, an end to end pipeline for joint channel and symbol estimation along with resource allocation can be explored.

\bibliographystyle{ieeetr}
\bibliography{main}

\begin{thebibliography}{10}

\bibitem{bhattacharya2024optimalpowerallocationtime}
S.~Bhattacharya, K.~Rajabalifardi, M.~A. Mohsin, R.~Pote, and J.~M. Cioffi, ``Optimal power allocation and time sharing in low rank multi-carrier wi-fi channels,'' 2024.

\bibitem{book}
J.~Cioffi, ``Data transmission theory,'' vol.~40, no.~4, pp.~49--60, 2024.

\bibitem{yin2006ofdma}
H.~Yin and S.~Alamouti, ``Ofdma: A broadband wireless access technology,'' in {\em 2006 IEEE sarnoff symposium}, pp.~1--4, IEEE, 2006.

\bibitem{noma6}
Y.~Saito, Y.~Kishiyama, A.~Benjebbour, T.~Nakamura, A.~Li, and K.~Higuchi, ``Non-orthogonal multiple access (noma) for cellular future radio access,'' in {\em 2013 IEEE 77th Vehicular Technology Conference (VTC Spring)}, pp.~1--5, 2013.

\bibitem{tse_viswanath_2005}
D.~Tse and P.~Viswanath, {\em Fundamentals of Wireless Communication}.
\newblock Cambridge, U.K.: Cambridge University Press, 2005.

\bibitem{6692652}
Y.~Saito, Y.~Kishiyama, A.~Benjebbour, T.~Nakamura, A.~Li, and K.~Higuchi, ``Non-orthogonal multiple access (noma) for cellular future radio access,'' in {\em 2013 IEEE 77th Vehicular Technology Conference (VTC Spring)}, pp.~1--5, 2013.

\bibitem{noma7}
M.~Zeng, A.~Yadav, O.~A. Dobre, G.~I. Tsiropoulos, and H.~V. Poor, ``On the sum rate of mimo-noma and mimo-oma systems,'' {\em IEEE Wireless Communications Letters}, vol.~6, no.~4, pp.~534--537, 2017.

\bibitem{wang2020drl}
X.~Wang, Y.~Zhang, R.~Shen, Y.~Xu, and F.-C. Zheng, ``Drl-based energy-efficient resource allocation frameworks for uplink noma systems,'' {\em IEEE Internet of Things Journal}, vol.~7, no.~8, pp.~7279--7294, 2020.

\bibitem{zhang2019energy}
Y.~Zhang, X.~Wang, and Y.~Xu, ``Energy-efficient resource allocation in uplink noma systems with deep reinforcement learning,'' in {\em 2019 11th international conference on wireless communications and signal processing (WCSP)}, pp.~1--6, IEEE, 2019.

\bibitem{mnih2016asynchronous}
V.~Mnih, A.~P. Badia, M.~Mirza, A.~Graves, T.~Lillicrap, T.~Harley, D.~Silver, and K.~Kavukcuoglu, ``{Asynchronous methods for deep reinforcement learning},'' in {\em Int. Conf. Mach. Learn.}, pp.~1928--1937, PMLR, 2016.

\bibitem{challita2019interference}
U.~Challita, W.~Saad, and C.~Bettstetter, ``{Interference management for cellular-connected UAVs: A deep reinforcement learning approach},'' {\em IEEE Trans. Wirel. Commun.}, vol.~18, no.~4, pp.~2125--2140, 2019.

\end{thebibliography}

\end{document}